\documentclass[10pt,twocolumn]{article} 
\usepackage{abstract}
\usepackage[utf8]{inputenc} 
\usepackage{amsxtra} 
\usepackage{amsthm}
\usepackage{url}
\usepackage{amssymb}
\usepackage{amsfonts}
\usepackage{graphicx}
\usepackage{rotating}
\usepackage{color}
\usepackage{epsfig}
\usepackage{subcaption} 
\usepackage{verbatim}
\usepackage[printonlyused]{acronym} 
\usepackage{setspace}
\usepackage{epstopdf}
\usepackage{authblk}
\usepackage{sectsty}
\usepackage[colorinlistoftodos]{todonotes}
\usepackage[normalem]{ulem}

\sectionfont{\fontsize{12}{15}\selectfont} 
\usepackage{lineno}
\usepackage{geometry} 
\renewcommand{\l}{\left}
\renewcommand{\r}{\right}

\newcommand{\Va}{V_{\text{age}}}
\newcommand{\Ve}{V_{\text{epi}}}
\newcommand{\Ven}{V_{\text{epi}}}
\newcommand{\Vmax}{V_{\text{max}}}

\title{Modelling the progression of atrial fibrillation:\\ A stochastic individual-based approach}

\author[1,4]{Eugene TY Chang}

\author[2,4]{Yen Ting Lin}

\author[2]{Tobias Galla}

\author[1]{Richard H Clayton}

\author[3]{Julie Eatock}

\affil[1]{Insigneo Institute for in-silico Medicine and Department of Computer Science, University of Sheffield, Sheffield, S1 4DP, UK}

\affil[2]{Theoretical Physics, School of Physics and Astronomy, The University of Manchester, Manchester M13 9PL, UK}

\affil[3]{Department of Computer Science, Brunel University London, Uxbridge, UB8 3PH, UK}

\affil[4]{These authors contributed equally.}

\begin{document}
\twocolumn[
  \begin{@twocolumnfalse}
	\maketitle
    \begin{abstract}
      We propose a stochastic individual-based model of the progression of atrial fibrillation (AF). The model operates at patient level over a lifetime and is based on elements of the physiology and biophysics of AF, making contact with existing mechanistic models. The outputs of the model are times when the patient is in normal rhythm and AF, and we carry out a population-level analysis of the statistics of disease progression. While the model is stylised at present and not directly predictive, future improvements are proposed to tighten the gap between existing mechanistic models of AF, and epidemiological data, with a view towards model-based personalised medicine.  \end{abstract}
      Keywords: atrial fibrillation, stochastic individual-based model, uncertainty quantification
      \\    \\ \\ \\
  \end{@twocolumnfalse}
]

\section{Introduction}

Atrial Fibrillation (AF) is an electrical heart disease characterised by irregular heart rhythm and absence of organised atrial activity, with a  prevalence of $1-2\%$ in the UK \cite{ncccc:2006}.
Incidence and progression of AF increases with age \cite{ncccc:2006}, and is correlated with hypertension, diabetes and other cardiovascular conditions \cite {Camm:2010, Nattel:2002, Potpara:2014}. Furthermore, AF significantly increases risk of stroke \cite{Lip:2010, Ntaios:2013, Wolf:1991}; thus, the importance of understanding this disease and of predicting its progression is undisputed. Aside from improved monitoring of patients and clinical intervention, computational modelling is becoming increasingly valuable in the medical sciences, in particular with a view towards personalised medicine \cite{Trayanova:2014}. A natural question is then to assess what is the most appropriate time and space scale over which to construct models of AF. Biophysically detailed models of cardiac cells and tissues can provide insights into the mechanisms that initiate and sustain AF \cite{Clayton:2011}, whereas population-level models, parametrised by available epidemiological data enable treatment pathways to be evaluated \cite{Lord:2013}. The most appropriate model will of course depend on the specific problem in hand.

Biophysically detailed models of cardiac cells and tissue are based on the physical laws governing the electro-mechanics of the heart. Simulating these models however is computationally expensive, and so only time scales of a few heartbeats in a single patient are accessible \cite{Trayanova:2014}. This constraint prevents a bottom-up simulation of progression of the disease in a single patient at a time scale longer than a few minutes, and so these models are not suitable for complete study of the self-reinforcing effect of AF (`AF begets AF' \cite{Wijffels:1995}) which occurs over days, weeks, and years. Constructing predictive models for the life-time progression of AF in individual patients based on the underlying biophysics is therefore difficult given current computational resources.

Population-level models can capture long time scales, but they require epidemiological data as an input to determine model parameters. Clinical data are available at medium time scales of several days, but these data are sparely sampled, and mostly collected from patients with implanted monitoring devices. These patients are more likely to have other cardiac-related co-morbidities, and so the sample is biased. In addition, the storage capacities of devices limit the temporal resolution of the data. To our knowledge, on a population-level, a quantitative model describing the progression of the disease at a long time scale is not yet available, and so the gap between acute mechanistic approaches and modelling on time scales of the human lifetime remains.

The aim of the present study is to make progress towards closing this gap and linking the small scale biophysical models with epidemiological models at a population level. Ultimately, one may hope that one day fully developed multi-scale models will be available, capable of predicting the progression of a given patient through the stages of AF over a time scale of years and starting from personalised parameters (e.g extent of atrial fibrosis obtained from cardiac imaging). Such models, if they can be constructed, would provide a key step towards computer-based predictive personalised medicine, and the ability to more accurately design individual treatment plans and to predict the cost-effectiveness of new interventions population-wide.

To work towards this long-term goal we propose a coarse-grained patient-level stochastic model, based on some elements of the current understanding of the mechanisms that spontaneously initiate, sustain, and terminate AF. The finest scale described by the model is the status of a patient with respect to the disease, i.e. a patient can at any one time be in an episode of AF, or in sinus rhythm (SR). The switching between these states is governed by two rates, describing initiation and termination of episodes respectively. Both of these rates are patient-specific in the model, and reflect genetic disposition, age and individual patient history. Most crucially, the model describes the effects of episode-based remodelling of the heart, making future episodes more likely. To capture this AF-begets-AF effect the model retains a transient memory of earlier episodes, and these affect the initiation and termination rates of later episodes. Many of these processes and the parameters describing them are---in principle---linked to the underlying biophysics. We stress clearly that establishing this link quantitatively is of course a major enterprise, and not within the scope of the present work. The spirit of our paper is very much to make a first step in this direction.

The remainder of the paper is structured as follows: In the next section we describe some of the main features of AF, both from the biophysical microscopic perspective and from the long-term epidemiological view. Our account is not comprehensive; we focus on features relevant to our modelling approach. In Section \ref{sec:modeldescription} we then present the actual mechanics of the model with a detailed quantitative description available in the Appendix for clarity. The outcome of our simulations is then described in Section \ref{sec:results}, before we summarise our model and discuss its potentials and limitations.

\begin{figure*}
\centering
\includegraphics[width=\textwidth]{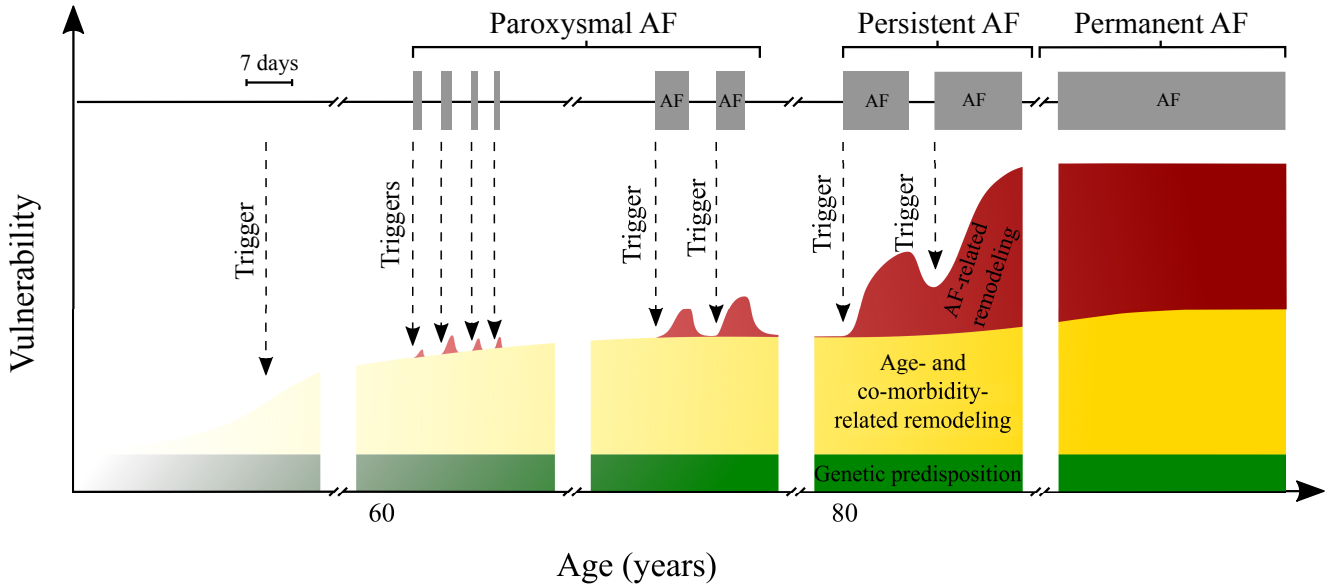}
\caption{\label{fig:AFcartoon}A framework of AF initiation, maintenance and progression, modified from \cite{Heijman:2014}. AF onset is dependent on vulnerability and triggers. Vulnerability is dependent on a constant genetic predisposition, time-varying age/co-morbidity-related remodelling and AF-related remodelling dependent on AF history. An increase in vulnerability  may allow initiation of AF episodes upon a trigger event occurring; over time, some patients progress on to paroxysmal, persistent and permanent AF as their AF vulnerability increases. Note the timescale for AF triggers and episodes is distinct from the lower axis and is expanded for visualisation. }
\end{figure*}
\section{Main features of AF relevant to the model}
\subsection{Biophysical processes}\label{sec:biophys}
We briefly summarise the biophysical phenomena contributing to the key processes described by the model: initiation and termination of episodes of AF, ignoring external interventions such as pacemakers, cardioversion, or effects of medication. The likelihood of initiation and termination of AF in the model is dependent upon remodelling of structural and and functional elements of the heart, which occur over multiple time scales \cite{Nattel:2008,Pandozi:2001,Schotten:2003}.

\subsubsection{Initiation of episodes of AF}
The transition from sinus rhythm (SR) to AF is characterised by a combination of a triggering event and an existing vulnerability of the atria to such triggers \cite{Schotten:2011}. This combination results in self-sustaining rotating waves of electrical activation, which act to suppress normal beats arising from the heart's natural pacemaker. Examples of triggering events are a rapid increase of the atrial rate, and ectopic electrical beats within the atria \cite{Chen:1999,Wakili:2011}. Increased vulnerability or triggers can be due to structural abnormalities in the atria such as fibrosis, hypertrophy or gap junctional lateralisation \cite{Everett:2000,Everett:2006, Thijssen:2001}, and specific electrical changes, which transiently modify the electrical behaviour of cardiac cells \cite{Nattel:2002, Wijffels:1995,Schotten:2011,Bosch:2002,Goette:1996,Yue:1999}. 

We adopted a coarse-grained picture of these detailed electrophysiological processes underpinning the initiation of episodes of AF. In our model the physiological state of the atria of a patient is described by a binary state: the patient can either be in AF or SR.
In our approach, we separate triggering events from the vulnerability of the atria. Trigger events are seen as external to the model, so we do not include changes in patient-specific behaviour or physiology affecting the frequency of triggers. For simplicity, we assume that triggering events occur at a constant rate, throughout the life time of a patient. The time between two trigger events is random and exponentially distributed. In our model the vulnerability then describes the probability that a given trigger event leads to an episode of AF: the higher the vulnerability of the atria, the greater the probability that a triggering event induces an AF episode.

We consider three factors contributing to a patient's vulnerability: (i) a genetic predisposition, (ii) an age and co-morbidity related factor,  and (iii) previous history of the patient. These factors were succinctly summarised by Heijman et al. \cite{Heijman:2014}. Based on this existing work we illustrate the different components contributing to vulnerability in  Fig.~\ref{fig:AFcartoon}. Genetic pre-disposition is constant and does not depend on age. The contribution related to age and co-morbidities increases with time; we assume implicit correlation of age to other co-morbidities such as pulmonary hypertension, hypertrophy etc, and do not therefore consider independent effects of specific co-morbidities in the present study. The third element---previous history of the patient---models the memory effects of AF. The vulnerability of a patient at any one time depends on the time spent in AF previously and on the number of episodes the individual has experienced so far, and when these occurred. 

Factors (ii) and (iii) capture the effects of physiological remodelling: Structural and electrical properties of the atria are modulated by risk factors including ageing and previous history of AF \cite{Nattel:2002,Schotten:2011}. All factors are patient-specific: different patients in the simulation can, in principle, have distinct parameter settings.

\subsubsection{Recovery}
The biophysical mechanisms leading to spontaneous termination of an episode of AF are much less studied and understood than those leading to the initiation of an episode. Our formulation is based on the hypothesis that the factors contributing to the rate with which episodes terminate are similar to those contributing to activation. Any factor leading to an increase in the initiation rate is taken to cause a reduction in the termination rate. Specifically, we assume one contributing factor reflecting age and co-morbidities. This term in the recovery rate decays (exponentially) as the patient ages, and does not depend on previous history of AF progression. A second component depends on previous history: the propensity to recover from an episode of AF is reduced by time spent previously in the AF state. 

\subsubsection{Acute clustering}
Recurrence of an AF episode shortly after an existing episode has been observed in clinical studies \cite{Burashnikov:2003}. Kaemmerer et al \cite{Kaemmerer:2001} suggest that the likelihood of recurrence is highest immediately after termination and decreases over time, and that a long episode is likely to be followed by multiple shorter episodes immediately after. In order to capture this effect we assume that activation rate is increased immediately after an episode and exponentially decays to baseline, which is independent of AF history. We assume the recovery rate follows a similar trend; however we already assumed that the history component of AF recovery rate transiently decreases as time in an episode increases. To capture the effect that long episodes are immediately followed by short episodes, we thus assume the recovery rate `overshoots' above baseline immediately after an episode terminates, then decays back to baseline.

\subsection{Progressive stages of AF}
Three stages have been broadly defined clinically in the progression of AF: (i) \emph{paroxysmal AF}, (ii) \emph{persistent AF}, and (iii) \emph{permanent AF}. This is briefly illustrated in the schematic of Fig.~\ref{fig:AFcartoon}).

In \emph{paroxysmal AF}, a fibrillation episode has a typical duration of minutes or hours, lasting up to $7$ days, with episodes spontaneously terminating. Inter-episode times can vary greatly from days to months \cite{Shehadeh:2002}, and inter-episode times will often reduce as the condition progresses, due to physiological remodelling \cite{Wijffels:1995,Yue:1999, Morillo:1995}. The next stage, \emph{persistent AF}, is indicated when the patient has AF episodes lasting $7$ days or more, which do not self terminate and require medical intervention such as cardioversion. A patient is in \emph{permanent AF} when the episodes have continued for more than $1$ year; typically interventions to terminate the fibrillation will fail. 

\begin{figure*}
\begin{center}
\begin{subfigure}[t]{0.45\textwidth}
\includegraphics[width=\textwidth]{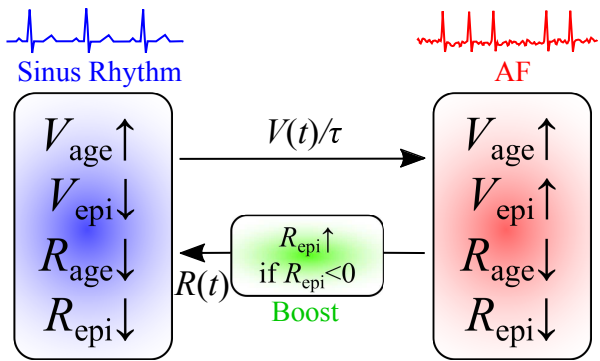}
\caption{Main elements governing the switching process between sinus rhythm and AF.}\label{fig:schematic1}
\end{subfigure}
\qquad
\begin{subfigure}[t]{0.4\textwidth}
\includegraphics[width=\textwidth]{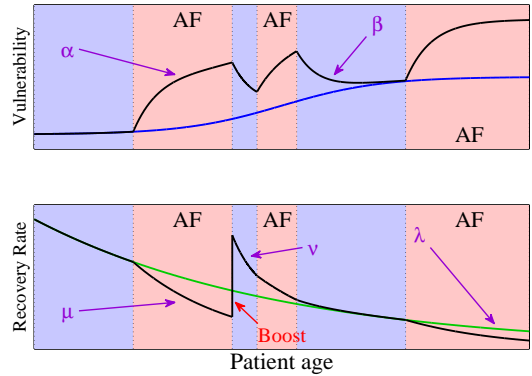}
\caption{Time courses of vulnerability and recovery rate during AF episodes and in sinus rhythm}\label{fig:schematic2}
\end{subfigure}
\caption{Illustration of the main mechanics of the model. Vulnerability of AF due to ageing, $\Va$, increases according to a sigmoid function, and age-based recovery rate $R_{0}$ decreases exponentially. A healthy patient triggers an AF episodes with rate $V(t)/\tau$ and terminates an AF episode with rate $R(t)$. In AF, episode-based remodelling $\Ven$ transiently increases vulnerability at a rate $\alpha$ and decreases the total recovery rate $R$ at rate $\mu$. When an episode terminates, the recovery rate `boosts' above baseline by a factor of $B$. Back in sinus rhythm, $\Ven$ decays to zero with rate $\beta$ and $R_{epi}$ decays with rate $\nu$ back to the age-based baseline $R_{0}$.}\label{fig:schematic}
\end{center}
\end{figure*}

We stress that although these stages are used to select appropriate treatment strategies, it is difficult to diagnose the stage any one AF patient is in clinically. Any such assessment is therefore approximate in practice. 

\subsection{Model Outline}
Our model describes a patient as either in AF or in sinus rhythm at any one time. A patient trajectory is therefore a series of points in time at which episodes begin and end. The starting point of the simulation is the birth of the patient, and all episodes are `detected' in the simulation. A real patient would not necessarily be diagnosed to have paroxysmal AF upon initial occurrence of an AF episode;  thus we use the following classification in our simulations. (i) A patient is considered to progress to paroxysmal AF immediately following an episode of at least $14.4$ minutes ($1\%$ of a day). (ii) They are said to be in persistent AF following an episode lasting $7$ days,  and (iii) progress to permanent AF from the point in time at which they enter an AF episode which continues until the end of the simulation at age $100$ years, without return to sinus rhythm.

\section{Description of the model}\label{sec:modeldescription}
\subsection{General setup}
The model describes the trajectory of individual patients throughout their life time. Patients can be simulated individually, and they do not interact. Time in the model is continuous, and at each point in time $t$ a patient is characterised by three attributes, their AF status $S(t)$, their vulnerability, $V(t)$, and their recovery rate $R(t)$. 

The AF status indicates if the patient is in AF at time $t$; if so we write $S(t)=1$, otherwise $S(t)=0$. As far as the occurrence of AF episodes is concerned a patient trajectory is fully characterised by the values $S(t)$ for all times $t$. Simulations start at age of zero years, this is labelled $t=0$. They end at age $100$ years ($t=100$).

The behaviour of $S(t)$ is determined by the activation and termination rates of AF episodes. If a patient is in sinus rhythm at time $t$ they can begin a new AF episode, this occurs with rate $V(t)/\tau$ in our simulation. The prefactor $\tau^{-1}$ indicates the frequency of trigger events. As described in earlier sections the initiation of episodes of AF occurs based on triggering events combined with existing vulnerability of the patient. The factor $\tau^{-1}$ reflects the trigger, and $V(t)$ stands for the vulnerability. 

If a patient is in an episode of AF at time $t$ ($S(t)=1$) they may return to sinus rhythm. This occurs with rate $R(t)$.

Mathematically speaking the model describes a binary process, in which $S(t)$ flips between the two states $S=0$ and $S=1$ at random times. The statistics of these transitions is governed by the time-dependent rates $V(t)/\tau$ and $R(t)$, both of which have units of inverse years. The underlying biophysical elements of AF initiation and termination we wish to describe (see Sec. \ref{sec:biophys}) are encapsulated in the detailed mathematical forms of these rates. It is important to keep in mind that these depend on the prior history of the patient, and so $V(t)$ and $R(t)$ are not set quantities, but random variables themselves. We will describe the mathematical forms of these rates next. In doing this we will first define the general mathematics, and not specify the numerical values of the various model parameters we are about to introduce. These will be discussed further below.

\subsection{Vulnerability $V(t)$}
 The vulnerability $V(t)$ is made up of three components; genetic predisposition, age and co-morbidity related vulnerability, and episode-induced vulnerability \cite{Heijman:2014}. We write this as

\begin{equation}
V_t = V_0 +\Va +\Ve.
\end{equation}
The quantity $V_0$ is the vulnerability associated with genetic predisposition, it is time-invariant and may vary from patient to patient. The second component, $\Va$, is the vulnerability associated with the age and with co-morbidities that may develop with time. In the present study we do not model co-morbidities independently or explicitly, and simply assume that $V_{\text{age}}$ depends primarily on age (i.e. other co-morbidities are implicitly correlated with age), but not on the prior AF-history of the subject. This type of vulnerability increases with time. More specifically we assume a sigmoidal function for the functional form of $\Va$,
\begin{equation}
\Va = \frac{V_1}{1+ \exp\l(-\frac{t-t_c}{t_d}\r)}.
\end{equation}
Age/co-morbidity-dependent vulnerability is low in the initial phases of a subject's trajectory (i.e. $t\approx 0$), and it then increases to plateau, $V_1$, at higher ages. The parameter $t_c$ describes the location of the turning point in this evolution, and $t_d$ characterises the width of the sigmoid, i.e. the duration of the transition between low vulnerability and the final plateau.

It remains to define $\Ve$ as the vulnerability associated with acute remodelling. The following piecewise process is adopted to model $\Ven$. When the patient is not in AF, i.e. at times when $S(t)=0$, we assume that $\Ven$ decays exponentially with rate $\beta$,
\begin{equation}
\frac{d}{dt} \Ven(t) = -\beta \Ven(t), \text{ when } S(t) = 0.\label{eq:evolutionofV1}
\end{equation}
When a patient is in AF (i.e., when $S(t)=1$), we assume that the vulnerability grows exponentially in time limited by a maximum vulnerability of $\Vmax$. Mathematically we write
\begin{equation}
\frac{d}{dt} \Ven(t) =\alpha \l[\Vmax - \Ven(t)\r], \text{ when } S(t) = 1 .
\label{eq:evolutionofV2}
\end{equation}
The model parameters $\alpha$ and $\beta$ are the relaxation rates of $\Ven$ back to $0$ when $S(t)=0$, or to maximum value when $S(t)=1$. 

We further assert that the vulnerability $V(t)$ is continuous: after a random switching event into AF or out of AF, the initial condition of $\Ven(t)$ in the following segment is set as the final value of $\Ven(t)$ right before the switching event.  A schematic diagram of the evolution of $V(t)$ is presented in Fig.~\ref{fig:schematic}.

\subsection{Recovery rate $R(t)$}
Next, we define the recovery rate $R(t)$ to consist of two components:
\renewcommand{\Re}{R_{\text{epi}}}
\newcommand{\Ren}{R_{\text{epi}}}
\newcommand{\Rage}{R_{\text{age}}}
\begin{equation}
R_t := \Rage(t) + \Re(t).
\end{equation}
The variable $\Rage(t)$ stands for a naturally declining ability to recover from an established episode. The decline is due to age, and it is taken to be exponential with rate $\lambda$,
\begin{equation}
\Rage(t) := R_0 e^{-\lambda t}. \label{eq:r0}
\end{equation}
The second component, $\Ren$, is the history-dependent recovery rate, which is known to be a function of preceding episodes of AF \cite{Wijffels:1995}. Similar to the above procedure for the vulnerability we define a piecewise process for $\Ren$. When the subject is not in AF, we assume
\begin{equation}
\frac{d}{dt} \Ren = - \nu \Ren(t) \text{ when } S(t) = 0, \label{eq:ren}
\end{equation}
which describes the recovery from future episodes of AF, and indicates an exponential decay of $\Ren$ with rate $\nu$ during times the patient is in SR.

When the patient is in the AF state ($S(t)=1$) we assume that the {\em total} recovery rate, $R(t)=\Rage(t)+\Ren(t)$ falls exponentially with rate $\mu$
\begin{equation}
\frac{d}{dt} R(t) = -\mu R(t), \text{ when } S(t) = 1. \label{eq:rtot}
\end{equation}
This different formulation is intended to capture the short term remodelling that takes place during an episode of AF, and acts to reduce the likelihood of termination. Crucially we assume $\mu>\lambda$, i.e. the drop in total recovery rate during AF is quicker than the drop of $\Rage(t)$ when the subject is in sinus rhythm. Typically the total recovery rate $R(t)$ will fall below the baseline rate $\Rage(t)$, hence $\Ren(t)=R(t)-\Rage(t)$ formally turns negative; we comment on this below. The dynamics illustrated in Fig.~\ref{fig:schematic}.

In order to capture the clustering of several short episodes frequently observed clinically after a long AF episode, we introduce an additional `boost' term into the dynamics of the recovery rate. Specifically we use the following procedure when an episode of AF terminates. Say the termination time is $t$. Prior to this moment the patient will have been in AF, and so $R(t)$ is governed by Eq.(\ref{eq:rtot}). The age-related contribution is $\Rage(t)=R_0e^{-\lambda t}$, see Eq.(\ref{eq:r0}). Together these equations define $\Ren=R-\Rage$ just before the end of the episode at $t$. After a long episode of AF this net rate $\Ren$ will often be negative. To capture the clustering we assume that immediately after the end of an episode the {\em total} recovery rate $R(t)$ overshoots and exceeds the natural background $\Rage(t)$. In order to implement this we introduce an instantaneous boost at the end of an episode: we change the sign of $\Ren(t)$ and we allow for an additional factor $B>0$ in its magnitude: $\Ren\rightarrow - B\times\Ren$. This is only implemented if $\Ren(t)<0$ just before the end of the episode, otherwise no reflection or boost are applied. We note that negative rates of $\Ren$ do not obviously have a direct meaning in isolation. In our model only the {\em total} recovery rate $R=\Rage+\Ren$, has a direct biological interpretation. This rate is always positive.

The different features of the dynamics are illustrated in Fig. \ref{fig:schematic}, where we also show stylised possible time courses of the vulnerability and recovery rate.

\subsection{Implementation}
The model constitutes a continuous-time stochastic process with discrete and continuous variables. During episodes and during times of sinus rhythm the rates $R(t)$ and $V(t)$ evolve deterministically according the dynamics specified by the ordinary differential equations above. The switching dynamics between $S=0$ and $S=1$ (SR and AF) is stochastic and governed by these rates. 
The transition rates evolve deterministically over time conditional on the current state of the subject. The model can be implemented relatively straightforwardly using a modified Gillespie algorithm \cite{Gillespie76,Gillespie77} to account for the time-dependent rates governing the statistics of the switching events into and out of AF. This is discussed further in the Appendix.

\begin{table*}
    \footnotesize
	\begin{tabular}{ | c | l | c | c| l |}
    \hline
    Parameter & Description &  Value & Unit & Comment/Reference \\ \hline\hline
    $1/\tau$ & Triggering rate &  $1$  &  1/year  & Basic simulation time unit\\ \hline
    $V_0$ & Gene. predis. vul. &  $ 10^{-5} $  &  n/a  & Set arbitrarily small\\ \hline
    $V_1$ & Max.~age/disease~vulnerability& 2920 & n/a & Paroxysmal AF patients have up to 8 episodes/day \cite{Hoffmann:2006} \\ \hline
	$V_{\rm max}$ & Max.~episode~vulnerability& $2\times v_1$& n/a &AF doubles VW in tissue \cite{Colman:2013}\\ \hline
	$t_c$ & Age/disease transition time& 70&  years& Mean incidence rate \cite[Table 1]{Haim:2015}\\ \hline
	$t_d$ & Age/disease transition width& 3&  n/a & Gives range $\pm 10$ years\\ \hline
	$\alpha$ & Episode vul.~saturation rate& 52 &  1/yr& AERP stabilises after one week \cite{Yu:1999}\\ \hline
	$\beta$ & Episode vul.~relaxation rate& 182.5&   1/yr& AERP recovers after two days \cite{Garratt:1999}\\ \hline
	$R_0$ & Peak recovery rate (r.r.) (of new born)& 0.5&  second & Min possible period of AF\\ \hline
$\lambda$ & Nat.~degradation rate of r.r.&  $1.2/50*\log(840)$ &  & Set so that patient has 14.4 min episode at age 50 \\ \hline
	$\mu$ & Degradation rate of total r.r. in AF & $\alpha$ &   & Follows $\alpha$\\ \hline
	$\nu$ & Relaxing rate of epi.~r.r. when Healthy & $\beta$ &   &Follows $\beta$\\ \hline
	$B$ & Boost factor of r.r. post AF& 1& n/a & 1: overshoot$\times$1; 0: no o/s\\ \hline 
	\end{tabular}
    \caption{The parameter set used in model simulations. Where possible, parameters were estimated from literature; certain parameters were set to reproduce incidence behaviours from epidemiological studies.}\label{tab:parameterset}
\end{table*}

\subsection{Model parameters}
The parameter set used in the present study is listed in Table \ref{tab:parameterset} and discussed below.  We stress that these are not definitive population-representative values, and that further exploration of the parameter space is necessary.

The external triggering rate $1/\tau$ scales the vulnerability $V(t)$ and has units of inverse time. We here use $\tau=1/yr$, and use years as our unit of time. In the following we will occasionally omit units in activation and recovery rates, as they are always understood to be per year. The numerical values of the  different components of $V(t)$ can then be set individually, as shown in Table \ref{tab:parameterset}. The parameters $V_0$, $V_1$ and $\Vmax$ are dimensionless quantities. We are focused on investigating effects of natural history and of AF-dependent remodelling, and thus set the vulnerability due to genetic predisposition $V_0$ to be very small at $10^{-5}$. Ageing is a key predictor of AF vulnerability \cite{Andrade:2014}, with incidence and prevalence rates increasing with age. For the vulnerability due to ageing/co-morbidity, $\Va(t)$, we set the maximum value $V_1$ as $365 \times 8 = 2920$ on the basis of Hoffmann et al \cite{Hoffmann:2006} who reported a median of $8$ episodes per day in a patient cohort with paroxysmal AF. For vulnerability due to AF history $\Ven$, we cite a recent computational study from Colman et al \cite{Colman:2013}, who predicted that chronic AF doubles the vulnerable time window (in which the atria is susceptible to formation of AF-inducing re-entrant circuits) compared to baseline; we take this to assume that $\Vmax = 2\times V_1$.

To identify the parameters of the sigmoid curve describing increase in vulnerability due to ageing, we looked at the incidence rates of AF in several population studies \cite{Haim:2015,Heeringa:2006,Lloyd-Jones:2004,Stewart:2001} and chose parameters $t_c = 70$ and $t_d = 3$, representing a sigmoid centred at 70 years of age, with a range of  approximately $10$ years.

The saturation and recovery rates for episode-related vulnerability were estimated from a series of experimental studies establishing the existence of AF-induced AF over timescales of days to weeks \cite{Garratt:1999, Wijffels:1995,Yue:1999}. They analysed the \emph{atrial effective refractory period} (AERP), a clinical biomarker of atrial vulnerability, over periods of artificially induced AF and looked at changes relative to baseline along with recovery rates post pacing. AERP is known to increase in the right atria with age \cite{Sankaranarayanan:2013} but shortens when in AF. From these studies, we set $\alpha$, the episode vulnerability saturation rate, to be $52$, corresponding to a time period of 1 week \cite{Yu:1999} and $\beta$, the episode vulnerability relaxation rate to $182.5$, corresponding to a maximum relaxation period of $2$ days \cite{Garratt:1999}. 

For the parameters governing the recovery function $R(t)$, we chose to estimate these based on intuition of the AF remodelling process. We set the peak recovery rate $\rho_0$ as $0.5$ seconds, which we assume to be the shortest possible duration of an AF episode - this corresponds to the fastest possible recovery time from AF of a (new born) patient in perfect health. The decay rate of the natural AF recovery function, $\lambda$ was derived on the basis that, from $\Rage=0.5$ at $t=0$, a typical episode of a 50 year old patient would be on the order of $14.4$ minutes in duration ($1\%$ of a day). This parameter would then be $\lambda = \log(840\times 2)/50$; in our simulation, we used a slightly larger value of $\lambda = 1.2\times\log(840)/50$.

Finally, for the episode-dependent recovery rate $\Ren(t)$, we assume that the degradation rate $\mu$ follows the saturation rate of episode-dependent vulnerability, $\alpha$, and that the relaxation rate $\nu$ similarly follows the corresponding relaxation, $\beta$ in the vulnerability function. For the boost factor $B$ (the overshoot above baseline immediately after an episode terminates), we set this to $1$, which exactly reflects the decreased amplitude in recovery rate from baseline due to the present episode of AF.

\begin{figure*}
\begin{center}
\includegraphics[width=0.95\textwidth]{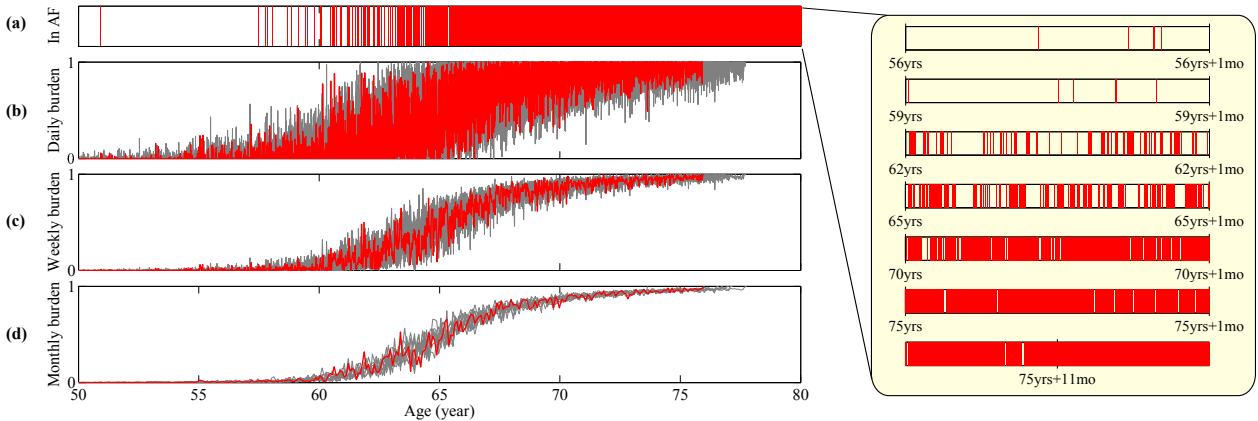}
\caption{Time series of the progression. A sample path is visualized by the red line, and from top to bottom: (a) Binary signal whether the patient is in an AF episode, (b) the fraction of the time in AF episodes per day, (c) the fraction of the time in AF episode per week, (d) the fraction of the time in AF episode per month. In (b-d) 9 other sample paths are plotted in grey lines in order to illustrate the distribution induced by intrinsic stochasticity. Inset: the progression of AF episode frequency and duration at time points in the simulation. }\label{fig:time_series}
\end{center}
\end{figure*}

\section{Model Outputs}\label{sec:results}
\subsection{General remarks}
The primary output of simulations of the model dynamics are patient trajectories. Each run generates a series of AF episodes over several decades; typically we see gradual progression of AF from paroxysmal to permanent. These synthetic data differ from existing clinical AF data: real-world AF ECGs are sampled at several kilohertz over short monitoring periods from hours to months, whilst our data simplifies the ECG in to sequences of binary AF/sinus rhythm episodes. The theoretical resolution of the model is dictated by the shortest episode generated by the model parameters.

We stress that the model parameters used are the same for all simulated patients. Each virtual patient has characteristics identical to the population average. Thus, our model does not capture heterogeneity in those parameters. A systematic analysis of the full parameter space is beyond the scope of the present work, but below we highlight a number of data which hint at the range of possible trajectories that this model generated. It should be noted that while the parameters as listed in Table \ref{tab:parameterset} are the same for all subjects in the model, different runs of the simulation will lead to different sequences of AF/SR episodes. This heterogeneity comes from the inherent randomness of the initiation and termination dynamics in the model.

Results reported below are from simulations of $5000$ independent patients. On each simulation run, the patient lives to age $100$ years, develops AF at some point and is found to enter permanent AF in their lifetime. This is not an accurate reflection of reality, of course, but this allows us to analyse the different patient trajectories as a consequence of both long- and short-term remodelling. 

\subsection{Patient trajectories and AF progression}
The top left panel of Fig.~\ref{fig:time_series} illustrates a sample time series plotted over a $30$-year time span. For any point in time the patient is either in AF or in sinus rhythm; episodes of AF are highlighted in red. In this example, an initial period of quiescence gives way to a sequence of AF episodes of increasing frequency and duration. As can be seen on the magnification on the right-hand side of the figure, where each time series represents a single month, there are many short and clustered episodes which may not be visible  over the longer time-scale view. These eventually become longer in duration until finally the patient remains in AF constantly until death (not shown).

The red (dark) curves in panels (b)-(d) of Fig.~\ref{fig:time_series} we show the daily, weekly, and monthly AF burden of the patient trajectory shown in panel (a). Following standard conventions \cite{Hoffmann:2006,Charitos:2014}, burden is here defined as the fraction of time spent in AF during a given observation window. To illustrate the variability of model outcomes panels (b)-(d) depict $9$ other sample paths (light grey). AF burden shows a monotonic increasing trend. Naturally the patient-to-patient variance reduces for longer observation windows (monthly versus daily). 

Using the parameter set in Table \ref{tab:parameterset}, we find that all patients develop paroxysmal AF at some point, and that they eventually enter permanent AF before the endpoint of the simulation at age $100$. 

\begin{figure}
\begin{center}
\includegraphics[width=0.48\textwidth]{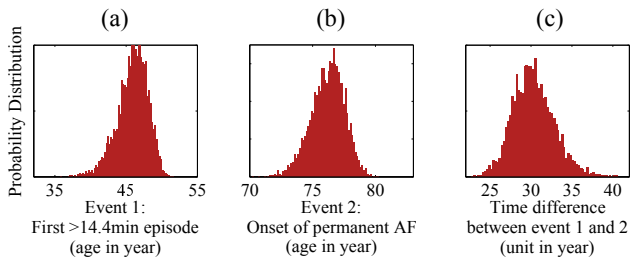}
\caption{Distributions of age at which paroxysmal AF sets it (panel (a)), age at which permanent AF sets in (panel (b)), and the time elapsed between these two events (panel (c)). Data were generated from $5000$ independent simulation runs of the model.}\label{fig:rough_progression}
\end{center}
\end{figure}

The model then allows us to simulate population statistics for the age at which patients develop paroxysmal AF (`event 1') and for the time at which they transition into permanent AF (`event 2'). These are shown in Fig.~\ref{fig:rough_progression} (panels (a) and (b)), along with statistics of the time that elapses between the onset of paroxysmal and permanent AF (panel (c)). It seems reasonable to assign a minimum threshold which would prompt a patient to present themselves for clinical diagnosis (assuming symptomatic AF). We set this threshold as an AF episode of duration greater than $14.4$ minutes ($1\%$ of a day). The model results show that the first episode of AF lasting longer than the threshold typically occurs when patients are aged around 44-48 years, which may be perceived as earlier than recorded in the literature \cite{Haim:2015,Heeringa:2006,Lloyd-Jones:2004,Stewart:2001}. Patients subsequently progress to permanent AF over a period of 28-32 years (panel (c)). A potential interpretation of these results is discussed in Section \ref{sec:discussion}.

\subsection{Behaviour after onset of paroxysmal AF}
We now turn to a discussion of model outputs in a form that may be clinically relevant. In our model we assume that the starting point of a patient's clinical trajectory is the time at which they develop paroxysmal AF. Given the intrinsic stochasticity, this time point will vary from patient to patient, as indicated in panel (a) of Fig.~\ref{fig:rough_progression}.

\begin{figure*}
\begin{center}
\includegraphics[width=.96\textwidth]{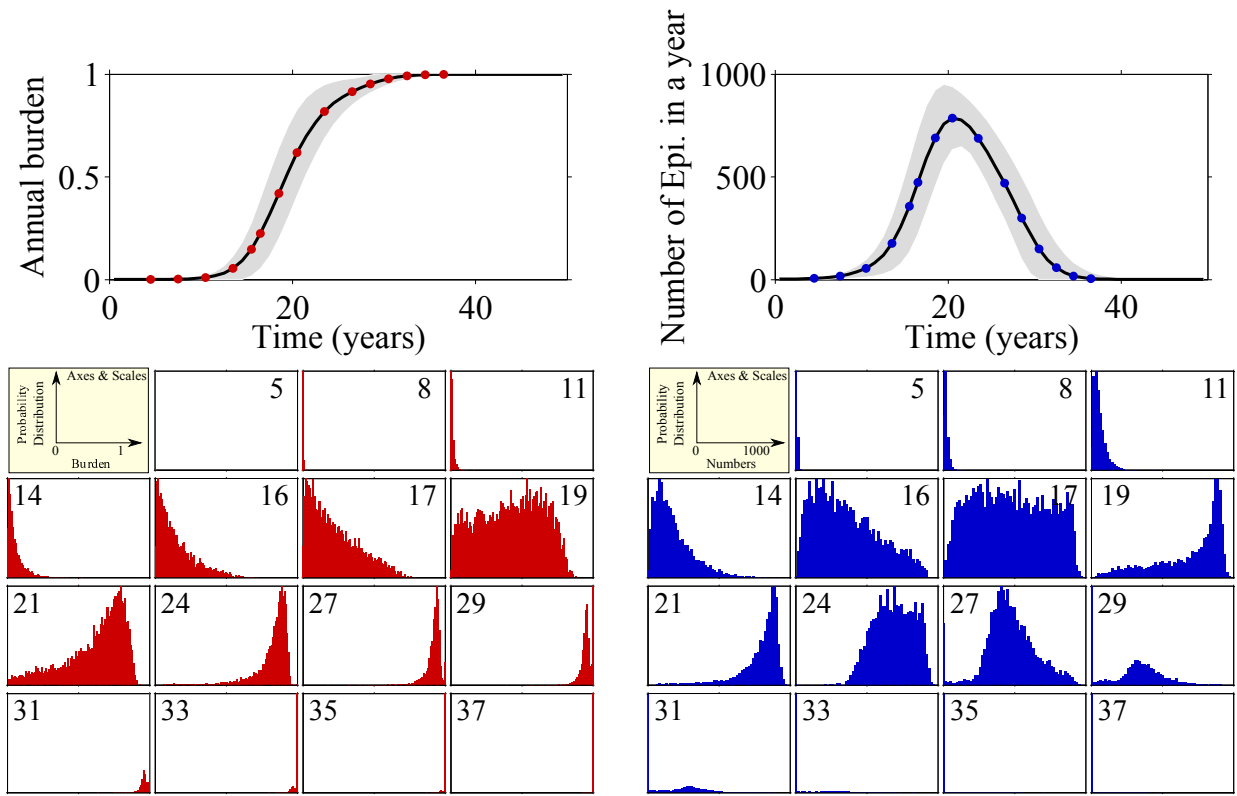}
\end{center}
\caption{{\bf Left:} Top panel shows annual AF burden after patient has entered paroxysmal AF. Horizontal axis indicates time elapsed since onset of paroxysmal AF. The mean of $5000$ simulation runs is indicated as solid line, grey area shows the standard deviation. Smaller panels at the lower left show distribution of burdens across the patient population at the indicated points in time (years after onset of paroxysmal AF). {\bf Right:} Analogous data for the number of AF episodes experienced in a moving window of $12$-months duration.}\label{fig:stat_after_14}
\end{figure*}

For any one patient we can use the time at which they enter paroxysmal AF as a reference time to follow their future progression. Data are shown in 
\ref{fig:stat_after_14}; the horizontal axes in the top panels of the figure indicate time after the onset of paroxysmal AF.

In the top left panel of Fig.~\ref{fig:stat_after_14}, annual AF burden is shown. The burden increases from near zero to one, representing the increase in the proportion of time spent in AF as the condition progresses. Each point on the curve represents the mean burden over $5000$ simulated patients. 

In the lower-left part of the figure we shown histograms of AF burden at the $15$ time points indicated in the top-left panel. The numbers in the smaller panels refer to the number of years since the onset of paroxysmal AF. At initial times after the onset of paroxysmal AF the burden shows a unimodal distribution, and over the following $17$ years the burden increases for some patients while remaining low for others producing a left skewed distribution. Around year $19$ there is a shift towards a higher burden and the distribution becomes more uniform, indicating that those patients that previously had low burden are now experiencing the effects of remodelling. By year $21 $these effects are evident in even more patients and the distribution becomes right skewed. Finally over years $24-37$ all patients experience the effects of remodelling and progress to permanent AF indicated by the distribution tending to a burden of $1$.

The right-hand side of Fig.~\ref{fig:stat_after_14} shows the total number of episodes per year (top panel) for the same dataset, along with detailed breakdowns of the population statistics in the lower right-hand panels. The number of episodes per year rises sharply and peaks at around 20 years after the onset of paroxysmal AF, and then gradually decreases. This reflects the observation that AF episodes begin intermittently, and become more frequent as structural remodelling increases vulnerability and decreases recovery rate. Consequently the episode durations lengthen and the inter-episode times shorten, so the number of episodes is reduced though the time spent in AF (AF burden) increases (left-hand panels of Fig.~\ref{fig:stat_after_14}). 

Initially the number of episodes experienced per year follows a unimodal distribution, with a peak at low numbers. The pace with which AF progresses in different virtual patients varies considerably; some will still have infrequent episodes so a low number of episodes per year, others will have experienced a moderate degree of re-modelling and will have many very frequent episodes. A third group will have experienced substantial structural re-modelling and so they have fewer episodes of longer duration. This leads to the broad distribution at around $17$ years after the onset of paroxysmal AF. Eventually all patients will progress to permanent AF and therefore have fewer episodes ($19$-$27$ years after threshold). Finally AF is permanent, i.e. there is one continuous episode, and then the centre of the distribution moves towards $1$ single episode per year.

\subsection{Behaviour prior to onset of permanent AF}
\begin{figure*}
\begin{center}
\includegraphics[width=0.96\textwidth]{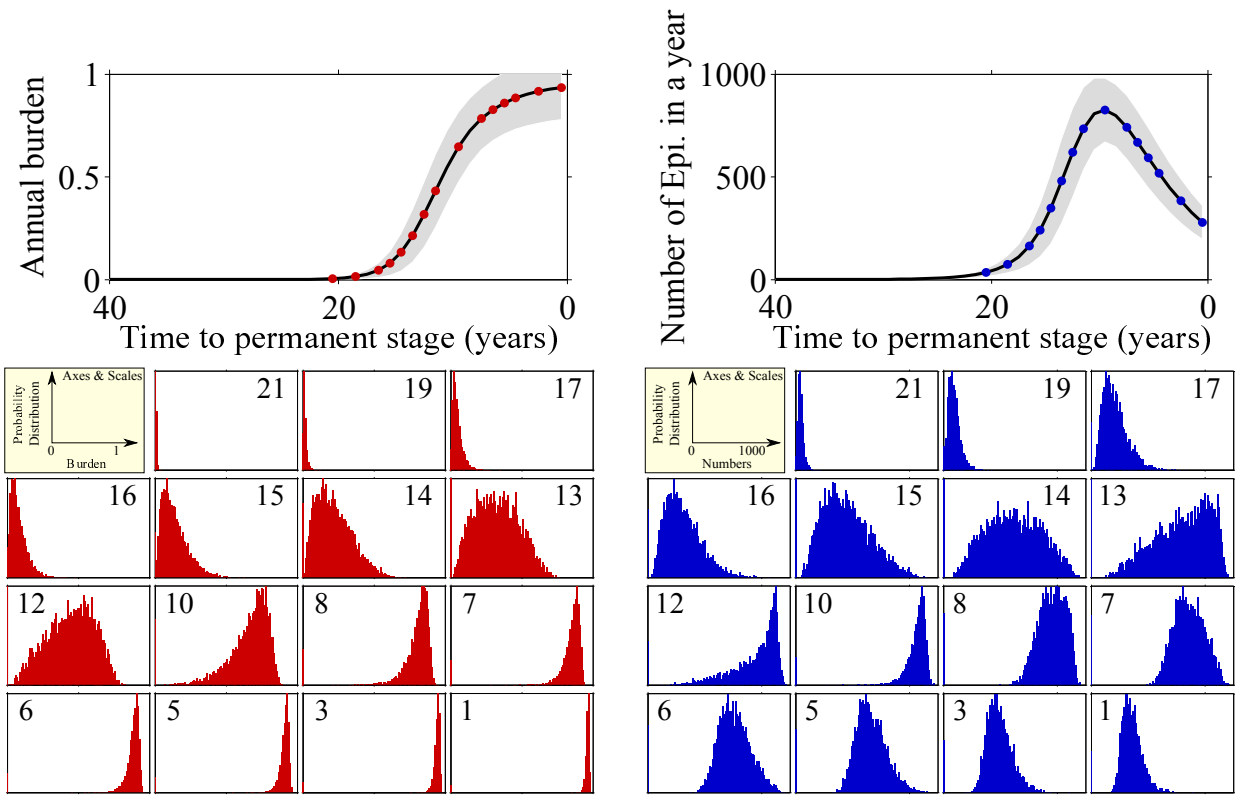}
\end{center}
\caption{(Left panel) Burden before patient goes into the permanent episode. (Right panel) Number of episodes before the patient goes into the permanent episode. In both panels we present the mean of the random variables (solid black) and mean $\pm$ 1 standard deviation (grey area) at the top, and at the bottom we measure the distributions of the random variables at 15 sampled time $t=\l [21,19,17,16,15,14,13,12,10,8,7,6,5,3,1\r]$ years before the permanent episode.}\label{fig:Burden_before_permanent}
\end{figure*}

We now change perspective and look at patient progression relative to the point in time at which they develop permanent AF. As discussed, different patients can progress differently after the onset of paroxysmal AF, and there will be considerable variation of the time at which they enter permanent AF. To allow for the  identification of potential patterns that can determine progression to permanent AF, we have replotted the model outputs in Fig.~\ref{fig:Burden_before_permanent}. Time is now defined relative to the point at which permanent AF sets in, the horizontal axis in the top panels now depict time until the onset of permanent AF. The left-hand panel shows the annual AF burden, for the time leading up to the moment of transition to permanent AF, the panels on the right show data for the number of episodes experienced per year.

From the left-hand panels in Fig.~\ref{fig:Burden_before_permanent} we can see that the annual burden begins to increase very slowly until 16 years prior to transition to permanent AF. At that point, patients start to progress at different speeds causing the distribution to become more unimodal (years $14-12$ prior to transition). By $10$ years before transition the patients are experiencing fairly high burden, which continues to gradually increase until they enter permanent AF.

Examining the right-hand panels in \ref{fig:Burden_before_permanent} one sees that the number of episodes at $17$ years prior to transition has noticeably increased but the corresponding burden (left-hand panel) is still low. This indicates that there are many short episodes - and we know that these are particularly unlikely to be picked up clinically simply due to timing. The number of episodes continues to increase and at $14$ years prior to transition the distribution for the number of episodes becomes more unimodal and symmetrical (ignoring a peak at zero episodes, reflecting patients who have not yet entered AF). The distribution shows a wide range in the number of episodes, with the corresponding burden also starting to increase. This illustrates that some people are experiencing the effects of short-term remodelling, and having many clustered episodes. At $12$ years before transition we see that most patients have many episodes, but burden still follows a relatively normal distribution. This indicates that some patients are experiencing the effects of long-term re-modelling and the duration of the individual episodes are increasing, while other patients are only experiencing the effects of short-term re-modelling and although they experience many episodes they are of short duration. From $6$ years until the transition to permanent the number of episodes decreases as the duration increases, until eventually the burden is $1$ and the patient remains in AF.

\section{Discussion}\label{sec:discussion}
In this study we have presented an individual-based model of AF progression. The model operates at patient level and operates a coarse-grained description of their progression through the different stages of AF. The fundamental variable of the model is the current AF status of the patient, being either in AF or in sinus rhythm. The transitions between these states are governed by history-dependent activation and termination rates, respectively. These rates are constructed to retain some key features of the physiology and biophysics of AF; in particular activation and termination rates both include terms describing acute electrical and structural remodelling. If a patient has a history of AF they are more likely to initiate further episodes, and these have a reduced probability to terminate. Other factors we include are a baseline vulnerability and age-dependent factors. The components related to age include co-morbidities which may develop over time, although we do not model these in detail. Our model builds on the conceptual framework proposed by Heijman et al \cite{Heijman:2014} and we remark that this conceptual model discusses only AF initiation and maintenance, but does not explicitly detail AF termination.

The formulation of our model is based on reaction kinetics with transitions between states that are modulated by the activation and recovery functions. This type of approach has been widely used in other areas (for example to describe the state of ion channels in cardiac cells), and yields behaviours that are broadly consistent with experimental observation. In the case of our model, we argue that it is reasonable to assume that patients with AF will transition between SR and AF states, and that the transition rates are affected by age and remodelling. The functional form of our age/co-morbidity-related vulnerability is consistent with the incidence curves reported in clinical studies \cite{Haim:2015,Lloyd-Jones:2004}. Where possible, we have chosen parameters based on evidence from the literature (Table \ref{tab:parameterset}). However, in a real patient we would expect some parameters such as the triggering rate to be variable rather than fixed as in the current model. We have also implicitly included the effects of co-morbidities within our description of vulnerability, which may not necessarily be age-correlated. Separating out the effects of co-morbidities in the vulnerability would allow us to investigate the progression of lone AF compared to other patient groups \cite{Jahangir:2007}. We expect that future studies will address these extensions of the model.

From the model outputs, we have analysed the typical disease progression of a virtual patient generated, and we have characterised the statistics derived from an idealised cohort of this population-averaged patient. In particular we report data relating to AF burden and the number of episodes experienced per year from the onset of paroxysmal AF forward. Secondly we have studied population statistics during the build-up to developing permanent AF. We did not study persistent AF in detail within the present work, other than to define the diagnosis threshold; the ability to predict progression to persistent AF would be of further clinical interest, given the difficulty of making accurate diagnoses from limited clinical monitoring windows.

The results from the model appear to show that the first episode with a duration greater than $14.4$ minutes, our diagnosis threshold for onset of paroxysmal AF, starts earlier in the patient's life than reported in the literature \cite{Haim:2015,Lloyd-Jones:2004}. However, it must be noted that the model records all events, while in reality these events may be asymptomatic, and therefore remain undiagnosed and therefore undocumented for many years \cite{Baturova:2014}. As a consequence of the recording of asymptomatic events prior to clinical diagnosis, the time taken to transition to permanent AF (Fig.~\ref{fig:rough_progression}(c)) may appear to be overly long. The follow-up period in published studies on the transition to permanent AF (e.g. \cite{Pappone:2008, Nieuwlaat:2008}) are of a duration of between $1$ and $5$ years or up to 30 years in studies of lone-AF (e.g. \cite{Jahangir:2007}), and include patients in both paroxysmal and persistent states at recruitment, so direct comparison with the present model result is challenging. However if we consider the age of transition of the patient to permanent AF, see Fig.~\ref{fig:rough_progression}(b), this does reflect findings in the literature that age is an independent predictor of chronic AF (both persistent and permanent) and that the risk associated with age increases beyond the age of 74 \cite{Lip:2010}. Additionally, this model takes no account of any medical interventions where the transition to permanent AF may be affected by different treatment strategies. Combining these results with models, such as Lord et al. \cite{Lord:2013}, that capture the longer-term impacts of medical interventions on a healthcare system may provide insights into when different treatment strategies are best applied along the progression trajectory. 

We stress that the model as it stands is of a stylised nature. While model parameters are chosen in agreement with existing clinical literature, we have made no attempt at validating the model outcome against real-world patient data. We do believe however that models of the type we propose here are placed somewhere between detailed mechanistic approaches and large-scale epidemiological studies, provide a promising way forward in integrating modelling of AF. We also note that analogous approaches are indeed possible for other diseases which progress through different stages \cite{Therneau12,Nowakbook,KeelingRohani}.

Now that a basic model is in place future work may focus on extending and refining it. Elements to incorporate in the future may include a more detailed breakdown of AF triggers, patient vulnerability and the recovery rate. This would ultimately have to be informed by the outputs of mechanistic models. Further important aspects include patient-to-patient variation of the underlying parameters, and ultimately the modelling of intervention techniques such as rhythm or rate control \cite{Roy:2008} At the other end, comparison and validation against long-term clinical cohort data would be desirable.

If progress can be made along these lines, one may hope that the intermediate modelling approach we propose here will ultimately establish a link between small-scale mechanistic models and large-scale epidemiological data. Properly validated individual-based models would ultimately contribute to the development of computer-based personalised medicine, and targeted pharmacological or surgical intervention.
\section*{Authors' contribution}
All authors designed the model, the study and the analysis. ETYC and YTL carried out the simulations and data analysis. All authors contributed to the writing of the final manuscript. All authors gave final approval for submission.
\section*{Competing interests}
We have no competing interests.

\section*{Funding}
We acknowledge funding by the Engineering and Physical Sciences Research Council EPSRC (UK), grant reference EP/K037145/1.

\appendix
\section{Further details of the simulation algorithm}
Simulation of the model generates sample paths, characterised by an alternating series of episode initiation and termination times. We here briefly describe  how these are obtained.

\subsection{Activation of episodes}
Assume that the individual is in sinus rhythm at time $t$, i.e. $S(t)=0$. Its current vulnerability and its recovery rate are known from the subject's prior history. The next step of the simulation is now to find the time $\Delta t$ until the onset of the next episode. The statistics of $\Delta t$ are governed by the time evolution of $V$ from time $t$ forward.

The subject is not in AF at time $t$, and so until the next episode sets in the time-dependence of vulnerability is given by ($t'>t$)
\begin{equation}
V(t')=V_0+\Va(t')+\Ven(t').
\end{equation}
The waiting time until the next episode, $\Delta t$, is then governed by the following cumulative probability distribution
\begin{equation}\label{eq:help}
\mbox{Pr}(\Delta t\leq T)=1-\exp\left(-\int_t^{t+T}dt' \frac{V(t')}{\tau}\right)
\end{equation}
Drawing $\Delta t$ from this distribution is non-trivial, and so we adopt a trick. While the onset of an episode in the clinic can obviously not be attributed to one unique the component to $V$ (i.e. either to $V_0$, $\Va$ or $\Ven$) we formally treat these as separate events in the simulation. I.e. we draw on waiting time, $\Delta t_1$ to an episode triggered with constant rate $V_0/\tau$, another waiting time $\Delta t_2$ generated using $\Va/\tau$, and another ($\Delta t_3$) based on $\Ven/\tau$. How this is done in practice is explained below. We then use the shortest of these three times, $\Delta t=\mbox{min}\{\Delta t_1, \Delta t_2, \Delta t_3\}$. This faithfully reproduces the statistics governed by the total rate, i.e. that of  Eq.~(\ref{eq:help}).

Simulation time is then advanced to $t+\Delta t$, the recovery rates are forward integrated, and an episode is activated at $t+\Delta t$ (i.e. we set $S=1$). 

It remains to explain how to generate waiting times until events triggered by the above three rates. 

Given that $V_0/\tau$ is constant the distribution of waiting times for events follows an exponential distribution with parameter $V_0/\tau$. This can be sampled directly setting $\Delta t_1=-\frac{\tau}{V_0}\ln~U_1$, where $U_1$ is a random variable drawn uniformly from the interval $(0,1]$.

To construct $\Delta t_2$, one plugs in $\Va(t')=V_1/ [\tau \l({1+\exp\left[-(t'-t_c)/t_d\r]}\r)]$. Integration gives

\begin{align}
{}&\mbox{Pr}(\Delta t_2\leq T) \nonumber\\
 {}&=1-\exp\l[- \frac{V_1 t_d}{\tau} \log \l(\frac{e^{(t+T)/t_d}+e^{t_c/t_d}}{e^{t/t_d}+e^{t_c/t_d}}\r)\r].
\label{eq:jerrymouse}
\end{align}
A sample of $\Delta t_2$ can then be obtained by drawing a uniform random number $U_2$ from the unit interval, and by solving 
\begin{equation}
\exp\l[-\frac{V_1 t_d}{\tau} \log \l(\frac{e^{(t+\Delta t_2)/t_d}+e^{t_c/t_d}}{e^{t/t_d}+e^{t_c/t_d}}\r)\r]=U_2.
\end{equation}
This can be done in explicit form.

Finally we construct $\Delta t_3$, using the same idea. We have $\Ven(t')=\l[V(t)-V_1(t) - V_0 \r]\exp\left(-\beta(t'-t)\right)$ using Eq.~(\ref{eq:evolutionofV1}), and keeping in mind that the subject is not in AF. After integration we see that $\Delta t_3$ can be obtained from setting
\begin{equation}
	  \exp\l[-\frac{V(t)-V_1(t)-V_0}{\tau\beta} \l(e^{-\beta t} -e^{-\beta \l(t + \Delta t_3\r)}\r)\r]=U_3.
\end{equation}
Again, this can be inverted in explicit form when 
\begin{equation}
U_3> \exp \l[-\frac{V(t)-V_1(t)-V_0}{\tau\beta} e^{-\beta t}\r],
\end{equation}
and we define $\Delta t_3 \rightarrow \infty$ otherwise.

\subsection{Termination of episodes}
Now assume the subject is in AF at time $t$. The algorithm proceeds by drawing a time $\Delta t$ to recovery. The recovery rate for $t'>t$ follows the exponential law $R(t')=R(t)e^{-\mu(t'-t)}$ during an episode, and so we can determine $\Delta t<\infty$ from
\begin{equation}
	  \exp\l[\frac{-R(t)}{\mu} \l(e^{-\mu t} -e^{-\mu \l(t+\Delta t\r)}\r)\r]=U,
\end{equation}
when 
\begin{equation}
 U>\exp\l[\frac{-R(t)}{\mu} e^{-\mu t}\r],
\end{equation}
and $\Delta t \rightarrow \infty$ otherwise. 
with a uniform random number $U$ drawn from the unit interval.

Once $\Delta t$ is drawn, activation and recovery rates are integrated forward until $t+\Delta t$, and the episode is terminated ($S(t+\Delta t)=0$). The algorithm then proceeds as described in the previous section.

\end{document}